\documentclass{article}
\usepackage{spconf,amsmath,graphicx}

\usepackage{enumitem}
\setlist{nosep, leftmargin=14pt}

\usepackage{mwe} 

\usepackage{booktabs}
\usepackage{bm}
\usepackage[super]{nth}
\usepackage{amssymb}
\usepackage[symbol]{footmisc}
\usepackage{url}
\title{Beyond the Monitor: Mixed Reality Visualization and Multimodal AI for Enhanced Digital Pathology Workflow}

\name{
    Jai Prakash Veerla$^{\star, 1, 3}$ \qquad
    Partha Sai Guttikonda$^{1, 3}$ \qquad
    Helen H. Shang$^{2}$
}

\secondlinename{
    Mohammad S. Nasr$^{1,3}$ \qquad
    Cesar Torres$^{\ddagger, 1}$ \qquad
    Jacob M. Luber$^{\dagger, 1, 3}$
}

\address{
    $^{1}$ Department of Computer Science and Engineering, University of Texas at Arlington\\
    $^{2}$ Department of Medicine Division of Hematology-Oncology, UCLA\\
    $^{3}$ Center for Innovation in Health Informatics, University of Texas at Arlington
}

\begin{document}

\maketitle
    \def\thefootnote{$\star$}\footnotetext{email: \texttt{jxv6663@mavs.uta.edu}} 
    \def\thefootnote{$\ddagger$}\footnotetext{email: \texttt{cearto@uta.edu}}
    \def\thefootnote{$\dagger$}\footnotetext{email: \texttt{jacob.luber@uta.edu}} 

    \begin{abstract}
Pathologists diagnose cancer using gigapixel whole-slide images (WSIs), but the current digital workflow is fragmented. These multiscale datasets often exceed $100,000 \times 100,000$ pixels, yet standard 2D monitors restrict the field of view. This disparity forces constant panning and zooming, which increases cognitive load and disrupts diagnostic momentum. We introduce \textit{PathVis}, a mixed-reality platform for Apple Vision Pro that unifies this ecosystem into a single immersive environment. \textit{PathVis} replaces indirect mouse navigation with embodied interaction, utilizing eye gaze, natural hand gestures, and voice commands to explore gigapixel data. The system integrates multimodal AI agents to support computer-aided diagnosis: a content-based image retrieval engine spatially displays similar patient cases for side-by-side prognostic comparison, while a conversational assistant provides real-time interpretation. By merging immersive visualization with integrated AI capabilities, \textit{PathVis} shows promise in streamlining diagnostic workflows and mitigating the burden of context switching. This paper presents the system architecture and a preliminary qualitative evaluation demonstrating the platform's feasibility. The PathVis source code and a demo video are publicly available at: {\footnotesize\url{https://github.com/jaiprakash1824/Path_Vis}}.
\end{abstract}

    \section{Introduction}
\label{sec:intro}

\begin{figure}[htb]

\begin{minipage}[b]{1.0\linewidth}
  \centering
  \centerline{\includegraphics[width=8.5cm]{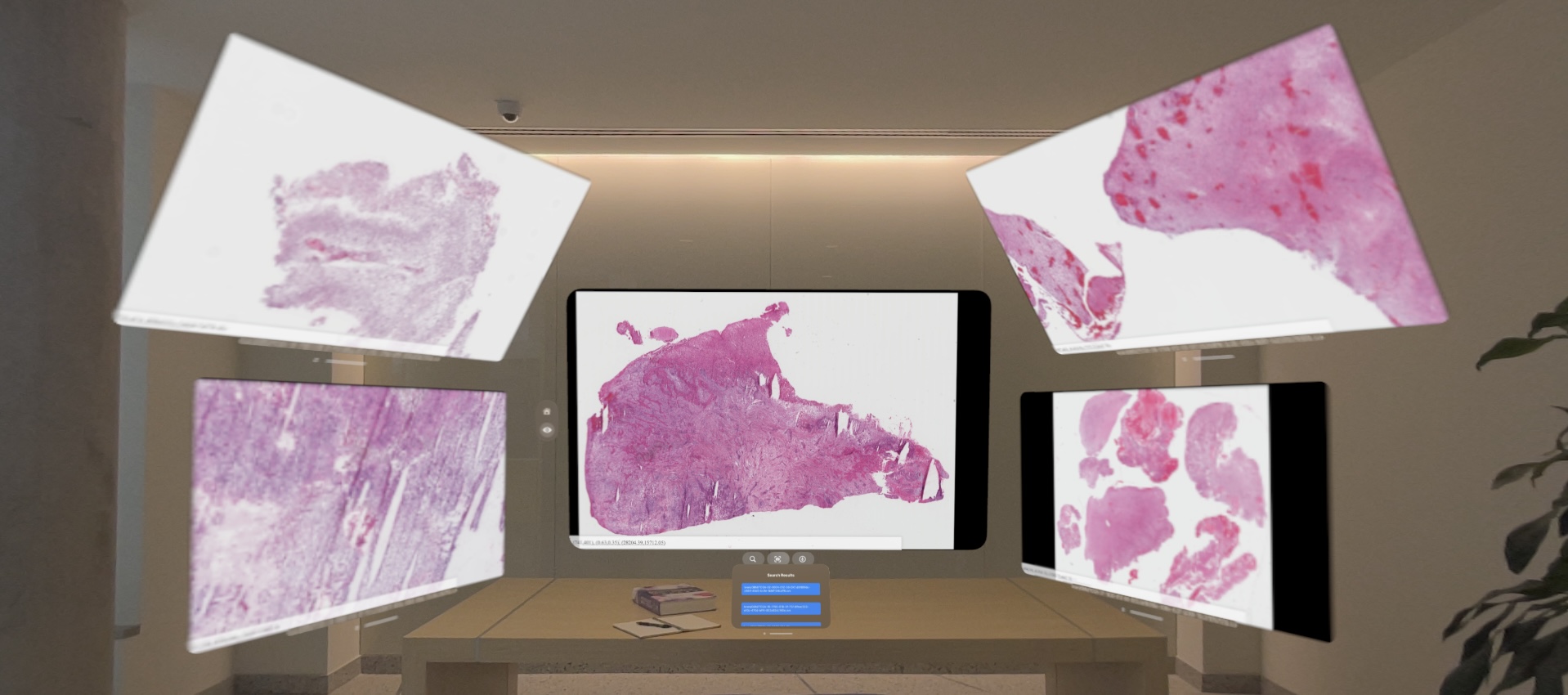}}
\end{minipage}
\caption{Screenshot of the \textit{PathVis} user interface operating in a mixed-reality environment on Apple Vision Pro. The central window displays the primary Whole Slide Image (WSI) under examination by the pathologist. The four surrounding windows show visually similar WSIs retrieved by the integrated AI-driven similar patient search feature, arranged spatially to facilitate direct comparison. Navigation within and between windows is performed using eye gaze and hand gestures.}
\label{fig:teaser}
\end{figure}

Digital pathology transforms cancer diagnosis by replacing physical glass slides with gigapixel whole-slide images (WSIs). These massive datasets often exceed $100,000 \times 100,000$ pixels~\cite{wang2012managing} and range from 5 to 30 gigabytes. While digital workflows facilitate collaboration~\cite{baxi2022digital}, analyzing these vast images on standard monitors presents significant \textbf{navigational bottlenecks}~\cite{10.1145/2834117}. Traditional viewers like QuPath~\cite{qupath} constrain the pathologist's field of view, necessitating constant panning and zooming to navigate Regions of Interest (ROIs). This fragmented interaction increases cognitive load and contributes to diagnostic fatigue~\cite{lan2023annotation, howbodiesmatter}. Furthermore, the current diagnostic ecosystem is often disjointed; utilizing advanced tools like Artificial Intelligence (AI) typically requires pathologists to switch contexts between the viewer and separate external applications, breaking their diagnostic flow.

\begin{figure*}[t]
    \centering
    \includegraphics[width=0.9\linewidth]{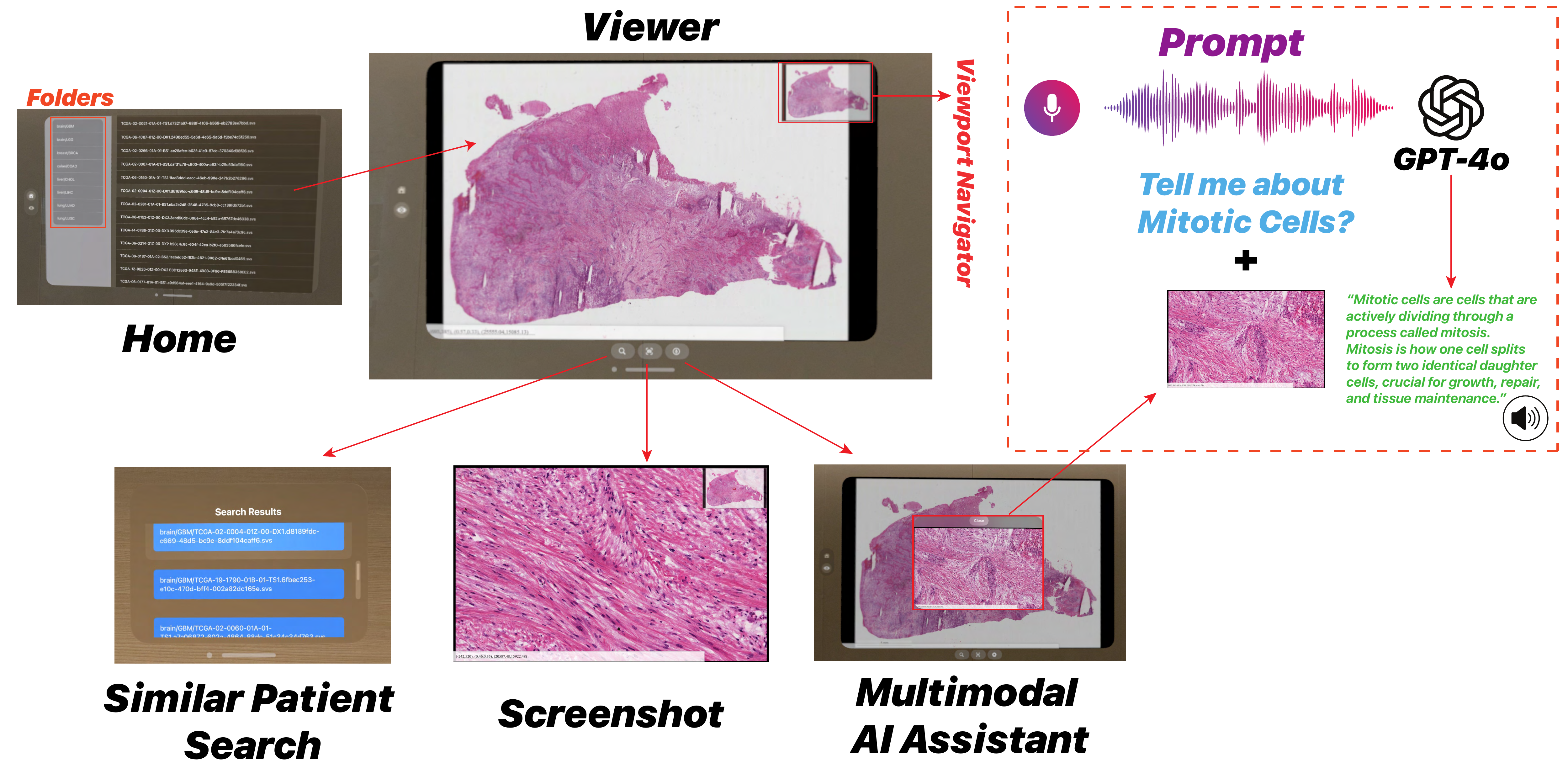}
    \caption{Overview of the PathVis system components and workflow.}
    \label{fig:overview}
\end{figure*}

Spatial computing leverages the \textbf{immersive capabilities of Mixed Reality (MR)}~\cite{mixed_reality} to offer a path beyond the physical monitor. Previous research utilized Augmented Reality (AR) and Virtual Reality (VR) for surgical planning and medical education~\cite{javaid2020virtual, farahani2016exploring}. 

In pathology, systems like SpatialVisVR~\cite{spatialvisvr} explore immersive environments for specialized multiplexed imaging. While systems like NaviPath~\cite{navipath} have explored human-AI collaboration for guided navigation on 2D displays, they remain constrained by the monitor's boundaries. However, a critical gap remains: seamlessly integrating these visualization tools with the \textbf{multimodal AI agents} necessary for modern diagnosis. While Content-Based Image Retrieval (CBIR) systems like Yottixel~\cite{yottixel} and Multimodal Large Language Models (LLMs) like OpenAI GPT-4o~\cite{openai2024gpt4o} and Google Gemini~\cite{gemini25pushingfrontier_2025} offer powerful analysis, they currently exist as isolated tools. Current setups force pathologists to switch contexts between the viewer and external AI applications, disrupting the diagnostic flow.

We introduce \textbf{\textit{PathVis}}, an MR platform for the Apple Vision Pro that unifies immersive visualization and AI assistance. Drawing on principles of embodied interaction~\cite{howbodiesmatter}, PathVis replaces indirect mouse control with natural hand gestures and eye gaze, leveraging physical actions to make engagement with complex digital information more intuitive. The system embeds AI directly into this workspace: an \textbf{AI-driven search} instantly retrieves similar patient cases for spatial side-by-side comparison (Fig.~\ref{fig:teaser}), while a \textbf{conversational AI assistant} provides real-time interpretation via voice commands (Fig.~\ref{fig:overview}). By merging the directness of physical slide review with integrated AI agents --- specifically a similar-patient retrieval engine and a conversational AI agent, PathVis aims to reduce cognitive friction and enhance diagnostic precision.

This paper introduces PathVis and makes the following contributions:
\begin{itemize}
    \item \textbf{Immersive WSI Visualization System:} The design and implementation of PathVis, a mixed-reality system for visualizing and interacting with gigapixel WSIs on Apple Vision Pro.
    \item \textbf{Multimodal Interaction Model:} An interaction model tailored for pathology WSI analysis in MR, mapping natural inputs (hand gestures, eye gaze, voice) for efficient navigation, zoom, and region-of-interest (ROI) selection.
    \item \textbf{Integrated Similar Patient Search:} An integrated visual search capability that identifies and displays similar patient cases within the mixed reality environment, enabling immediate side-by-side comparison to support decision making without disrupting workflow.
    \item \textbf{Conversational AI Assistant:} A voice-activated AI system embedded directly in the diagnostic workspace that provides real-time image interpretation support and responds to pathologist queries, enhancing analysis flexibility without requiring context switching.
\end{itemize}
    \section{Methods}
\label{sec:methods}
We developed PathVis as a mixed-reality platform that unifies high-performance gigapixel visualization with integrated AI tools to streamline the digital pathology workflow.

\subsection{System Architecture and WSI Streaming}
\label{subsec:architecture}
PathVis employs a client-server architecture designed to bridge the gap between massive Whole Slide Images (WSIs) and the mobile constraints of head-mounted displays. The backend, implemented in Python Flask and running on high-performance hardware (e.g., NVIDIA A100 GPUs or M2 Ultra), acts as the central orchestration engine. It processes raw WSI files often exceeding 10GB using the OpenSlide library~\cite{openslide} to generate multi-resolution image pyramids. To enable fluid interaction on the Apple Vision Pro, we utilize an efficient tile-based streaming protocol used by viewers like OpenSeadragon~\cite{openseadragon}. The VisionOS client application communicates via a REST API, requesting only the specific image tiles required for the user's current viewport and resolution level. This on-demand strategy, supported by server-side caching (SlideCache), ensures low latency and high color fidelity. It allows users to scale fluidly from tissue overviews ($\approx$5 mm) down to cellular details ($\approx$0.25 µm) without the latency of loading full datasets locally.

\subsection{Immersive Interaction Model}
\label{subsec:interaction}
\begin{figure}[t]
\centering\includegraphics[width=0.9\linewidth]{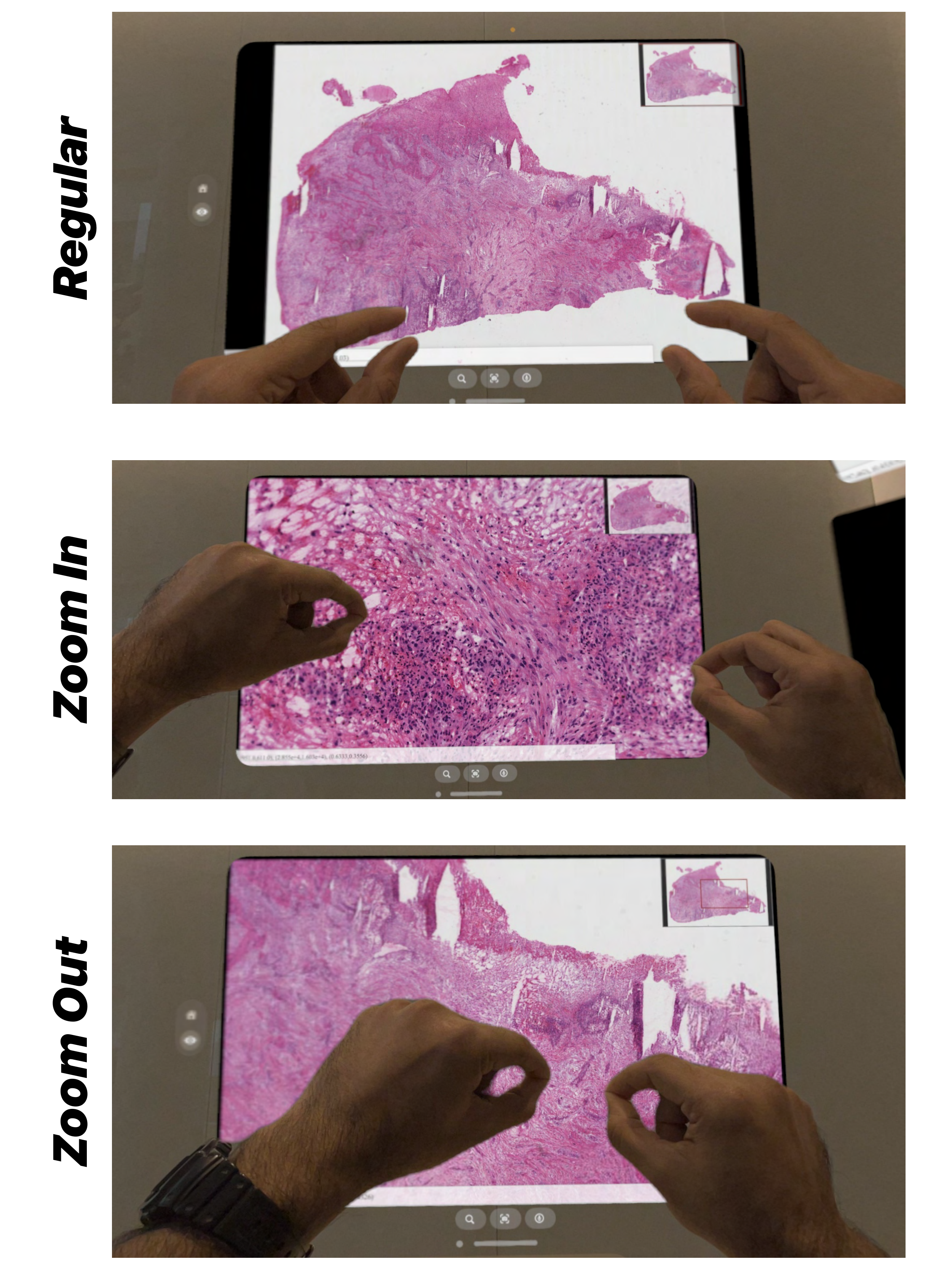}
\caption{Zoom interaction in PathVis using the two-handed pinch gesture. Moving hands apart zooms in (middle panel) for detailed examination, while moving them together zooms out (bottom panel).}
\label{fig:zoom}
\end{figure}

\begin{figure}[t]
     \centering
     \includegraphics[width=\linewidth]{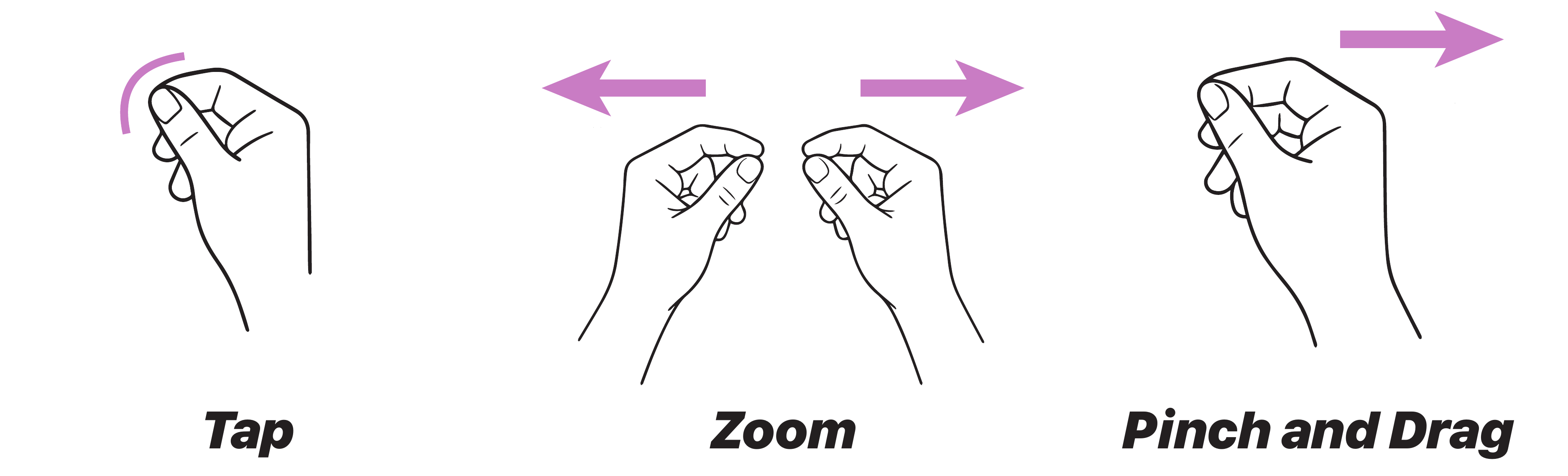} 
     \caption{Core hand gestures used for interaction in PathVis.}
     \label{fig:hands}
\end{figure}

We designed the interaction model to minimize cognitive friction by leveraging embodied actions rather than screen-based interactions. PathVis replaces the mouse with eye gaze as the primary pointing mechanism for targeting UI elements or tissue regions. Hand gestures provide direct manipulation control: a \textbf{\textit{tap}} gesture selects gaze-targeted items, a \textbf{\textit{pinch-and-drag}} gesture pans the WSI across the infinite virtual canvas, and a \textbf{\textit{two-handed pinch}} controls magnification (Fig.~\ref{fig:zoom}). This multimodal input scheme allows pathologists to navigate large datasets using natural physical movements (Fig.~\ref{fig:hands}), potentially reducing the ``mouse mileage" associated with traditional monitors. These gestures were explicitly selected to mimic the physical manipulation of glass slides on a microscope stage, leveraging the pathologist's existing muscle memory for intuitive control.

\subsection{Integrated AI Modules}
\label{subsec:ai_modules}
\textbf{\textit{Design Note:}} In this prototype, the assistant functions as a workflow co-pilot for information retrieval rather than a primary diagnostic device. We acknowledge that general-purpose LLMs require further domain-specific validation and fine-tuning to ensure clinical safety in diagnostic scenarios. 

A key contribution of PathVis is the architectural integration of distinct AI components directly into the mixed-reality workspace, reducing the need for context switching.

\subsubsection{Similar Patient Search}
\label{subsubsec:patient_search}
To support differential diagnosis, we embedded a Content-Based Image Retrieval (CBIR) engine. Users can trigger a search based on the current viewport or a specific Region of Interest (ROI). The system queries the Yottixel engine~\cite{yottixel} to identify the top-$k$ ($k=5$) morphologically similar cases from the NCI GDC database~\cite{nci}. Unlike traditional viewers that require opening new windows, PathVis displays these retrieved cases on virtual panels floating spatially around the central view (Fig.~\ref{fig:teaser}), facilitating immediate side-by-side prognostic comparison. We emphasize that this visual retrieval is designed strictly as a comparative reference aid for decision support and is not intended for primary diagnosis.

\subsubsection{Conversational AI Assistant}
\label{subsubsec:conversational_ai}
We integrated a voice-activated assistant using the OpenAI GPT-4o API~\cite{openai2024gpt4o} to demonstrate hands-free information retrieval (Fig.~\ref{fig:overview}). Pathologists can query the system regarding visual features, scoring protocols (e.g., Ki67 guidelines), or differential diagnoses. The assistant delivers audio responses directly within the environment, maintaining the user's visual focus on the slide.

    \section{Application Scenarios and Observations}
\label{sec:scenarios}

As an initial prototype, PathVis has not yet undergone formal, IRB-approved usability studies. The system was developed in consultation with a co-authoring oncologist to ground its features in clinical needs. The following scenarios illustrate the system's \textit{intended} applications, and the observations are based on informal feedback gathered from pathologists and other doctors during technical demonstrations (e.g., at CES) and throughout the development process.

\subsection{Fluid WSI Navigation and Exploration}
\label{subsec:scenario_nav}
Pathologists are accustomed to the tactile fluidity of moving physical glass slides, a directness that traditional mouse-based WSI navigation lacks. PathVis aims to restore this direct manipulation by replacing the mouse with gaze targeting and natural hand gestures (Figs.~\ref{fig:zoom} \&~\ref{fig:hands}).

\noindent\textbf{Observations:}
Informal feedback from co-authoring oncologist during demonstrations indicated that basic navigation is intuitive and mitigates the extensive mouse movement required by conventional 2D viewers. However, matching the speed and fine precision that pathologists have honed over decades of practiced microscope use remains a significant challenge. Observers noted that while gestures offer a potential reduction in repetitive strain, achieving expert-level efficiency for tasks like complex ROI outlining will require further refinement of gesture sensitivity and input mappings.

\subsection{Integrated AI for Clinical Decision Support}
\label{subsec:scenario_ai}
Diagnostic workflows often require ``context switching" to access external information, such as searching for reference cases (for differential diagnosis) or consulting guidelines(for quantitative tasks like Ki67/PD-L1 counting). PathVis integrates AI tools to provide this support directly within the immersive view.

\noindent\textbf{Observations:}
During demonstrations, the \textbf{Similar Patient Search} (Fig.~\ref{fig:teaser}) was frequently highlighted by pathologists as the most impactful feature. Pathologists confirmed that spatially arranging reference slides in the peripheral view without obscuring the primary WSI offers a significant advantage over the ``tab-switching" required on 2D monitors. The potential to have reference slides in the peripheral view, without obscuring the primary WSI, was seen as a significant advantage. 

Similarly, co-authoring oncologist noted that using the \textbf{Conversational AI Assistant} (Fig.~\ref{fig:overview}) for hands-free querying of scoring criteria could save time. However, the practical utility of these AI integrations is currently constrained by the latency and accuracy of the external APIs (Yottixel, OpenAI). Furthermore, managing multiple spatial information windows without causing visual overload remains a new and non-trivial design challenge for immersive clinical interfaces.
    \section{Discussion and Future Work}
\label{sec:discussion}

PathVis integrates immersive WSI visualization, natural multimodal interaction (eye gaze, hand gesture, voice), and integrated AI agents within a mixed-reality environment. This approach targets the cognitive load inherent in traditional screen-based navigation by offering an expansive view, intuitive controls, and readily accessible contextual AI support.

\textbf{Limitations:} As a framework implementation, this study primarily validates technical feasibility. We acknowledge that the pixel-level precision of eye-gaze navigation for fine-grained tasks (e.g., cellular annotation) remains unquantified in this pilot study. Additionally, the reliance on general-purpose LLMs introduces potential hallucination risks, which currently limits the system to a decision-support role pending further safety integration. Finally, the ergonomic impact of the headset's weight and form factor during prolonged clinical shifts remains unquantified; longitudinal studies are required to determine its suitability for routine primary diagnosis.

\textbf{Future Work:} Our immediate priority is rigorous validation through an IRB-approved comparative study. This will specifically quantify \textbf{Time-to-Diagnosis}, \textbf{System Usability Scale (SUS)}, and \textbf{Interaction Precision} compared to standard viewers. We also plan to formally measure network latency to ensure it meets clinical motion-to-photon thresholds. To address AI safety and data privacy, we plan to \textbf{train custom domain-specific models from scratch} and develop \textbf{specialized agentic frameworks} that leverage context-aware reasoning to ensure verifiable clinical outputs.
    \vspace{-0.75em}

\section{Conclusion}
\label{sec:conclusion}

Traditional digital pathology workflows often struggle with the immense scale and navigational challenges presented by gigapixel Whole Slide Images. In this paper, we introduced PathVis, a mixed-reality framework that combines large-scale spatial visualization, embodied interaction, and integrated AI agents for similar-patient search and conversational decision support. Through a detailed architectural overview and preliminary qualitative feedback, this work demonstrates the potential of PathVis to create fluid, context-rich diagnostic workflows that reduce the cognitive burden of tool fragmentation. While immediate future work focuses on rigorous quantitative validation and AI safety integration, PathVis represents a compelling step toward a future where immersive analytics and domain-specific agents empower pathologists with tools designed for more efficient and insightful analysis.
    \vspace{-1.5em}

\section{Acknowledgments}
\label{sec:acknowledgments}

 This work was supported by the University of Texas System Rising STARs Award (J.M.L) and the CPRIT First Time Faculty Award (J.M.L). Additionally, the research presented in this paper was supported by the National Science Foundation under Grant No. CNS: 2239646.

\section{Compliance with ethical standards}
This research study was conducted retrospectively using human subject data made available in open access. Ethical approval was  \textbf{not} required as confirmed by the license attached with the open access data.

    \bibliographystyle{IEEEbib}
    \bibliography{refs}

\end{document}